\documentclass{article}

\usepackage{arxiv}

\usepackage[utf8]{inputenc} 
\usepackage[T1]{fontenc}    
\usepackage{hyperref}       
\usepackage{url}            
\usepackage{booktabs}       
\usepackage{amsfonts}       
\usepackage{nicefrac}       
\usepackage{microtype}      
\usepackage{lipsum}
\usepackage{graphicx}
\usepackage{amssymb}
\usepackage{amsmath}
\usepackage{cleveref}
\usepackage{multirow}
\graphicspath{ {./images/} }

\title{FAN-Unet: Enhancing Unet with vision Fourier Analysis Block for Biomedical Image Segmentation}

\author{
 Jiashu Xu \\
  School of Science\\
  Harbin Institute of Technology, Shenzhen\\
  \texttt{jiashu.xu04@gmail.com} \\
}

\begin{document}
\maketitle
\begin{abstract}
Medical image segmentation is a critical aspect of modern medical research and clinical practice. Despite the remarkable performance of Convolutional Neural Networks (CNNs) in this domain, they inherently struggle to capture long-range dependencies within images. Transformers, on the other hand, are naturally adept at modeling global context but often face challenges in capturing local features effectively. Therefore, we presents FAN-UNet, a novel architecture that combines the strengths of Fourier Analysis Network (FAN)-based vision backbones and the U-Net architecture, effectively addressing the challenges of long-range dependency and periodicity modeling in biomedical image segmentation tasks. The proposed Vision-FAN layer integrates the FAN layer and self-attention mechanisms, leveraging Fourier analysis to enable the model to effectively capture both long-range dependencies and periodic relationships. Extensive experiments on various medical imaging datasets demonstrate that FAN-UNet achieves a favorable balance between model complexity and performance, validating its effectiveness and practicality for medical image segmentation tasks. 
\end{abstract}


\section{Introduction}
Modern medical research relies heavily on the support of various medical imaging technologies\cite{litjens2017survey}. Medical imaging aims to provide accurate visual representations of the structure and function of human tissues and organs, offering essential insights for healthcare professionals and researchers to explore normal and abnormal conditions in detail.Whether in cutting-edge laboratory studies or clinical disease diagnosis, the rich information provided by medical image analysis is crucial for scientific inference and diagnosis\cite{chen2022recent}. Furthermore, automatic medical image segmentation technology can assist doctors in faster pathological diagnosis, thereby improving patient care efficiency.

Due to their strong feature representation capabilities, Convolutional Neural Networks (CNNs) have been widely applied in the field of medical image segmentation and have achieved promising results\cite{shin2016deep}. Fully Convolutional Networks (FCNs) \cite{sun2021segmentation}, as an advanced variant of CNNs, enable pixel-level segmentation for images of arbitrary sizes. Building on this foundation, U-Net emerged as an innovative approach based on FCNs \cite{ronneberger2015u}, leveraging a symmetric encoder-decoder structure with skip connections to enhance the integration of contextual information effectively. Raj et al. \cite{raj2022enhanced} highlighted the effectiveness of U-Net-based methods for MRI data segmentation, while Safi et al. \cite{safi2023accurate} demonstrated the capability of CNNs to handle brain tumors of varying sizes, locations, and blurred boundaries. These findings collectively emphasize the robust performance of CNNs, particularly in medical image segmentation tasks such as brain tumor segmentation.  

Although CNN-based models possess excellent feature representation capabilities, they face inherent limitations in capturing long-range dependencies within images due to the locality of convolutional kernels. In contrast, Transformer models are naturally adept at capturing global context, which has sparked growing research interest in Transformer-based architectures for medical image segmentation. TransUnet\cite{chen2021transunet}, a pioneering Transformer-based model, first utilizes the Vision Transformer (ViT) for feature extraction during the encoding stage and then employs CNNs for decoding, demonstrating strong capabilities in capturing global information. TransFuse\cite{zhang2021transfuse} combines ViT and CNN in a parallel architecture, capturing both local and global features simultaneously. In addition, Swin-UNet\cite{cao2022swin} integrates the Swin Transformer into a U-shaped architecture, introducing the first purely Transformer-based U-Net model.These advances underscore the potential of Transformer-based methods in addressing the limitations of CNNs and enhancing segmentation performance.

Beneath the apparent success of current neural networks lies a critical issue: their difficulty in modeling periodicity from data. Existing neural network architectures, including MLPs\cite{rosenblatt1958perceptron}, KAN\cite{liu2024kan}, and Transformers\cite{vaswani2017attention}, struggle to fit periodic functions, even for simple cases such as sine waves\cite{dong2024fan}. While these models exhibit strong interpolation capabilities within the domain of training data, they often falter when faced with the challenge of extrapolation, particularly in out-of-distribution (OOD) scenarios. As a result, their generalization performance largely depends on the scale and diversity of the training data, rather than being driven by an inherent understanding of periodic principles.In the context of medical imaging data, pathological features often exhibit consistent periodic patterns\cite{thompson2014quasi}. Therefore, the ability to learn and model periodic principles is of paramount importance for medical image analysis. Such capability not only enhances the robustness of these models but also improves their effectiveness in identifying and interpreting repetitive patterns commonly seen in medical images.

In recent years, Fourier Analysis Network (FAN)\cite{dong2024fan} has emerged as a groundbreaking neural network framework, demonstrating exceptional capabilities in periodic modeling. Built upon Fourier analysis, FAN leverages Fourier series to explicitly encode periodic patterns within neural networks, offering a principled approach to modeling periodicity directly from data. Compared to traditional MLPs and the newly proposed KAN, FAN not only maintains efficiency but also excels in handling periodic features, showcasing superior performance in this domain.

In medical image segmentation tasks, although target features often exhibit high similarity, their shapes and sizes tend to vary significantly. Effectively capturing these variations is crucial for model performance. The U-Net-based architecture provides a robust framework for feature extraction, enabling the FAN layer to perform periodic modeling more effectively. Furthermore, integrating the FAN layer with self-attention mechanisms allows the model to simultaneously excel in both periodic modeling and long-range dependency modeling, enhancing its overall capability in handling complex medical imaging scenarios.

In this paper, we propose FAN UNet, a novel architecture that combines visual FAN to overcome the inherent challenges of long-term dependencies and periodicity in biomedical image segmentation task models. The Vision-FAN layer combines the strengths of the FAN layer and self-attention mechanisms, leveraging Fourier analysis to enable the model to effectively capture both long-range dependencies and periodic relationships. Extensive experiments on various medical imaging datasets demonstrate that FAN-UNet achieves a favorable balance between model complexity and performance, validating its effectiveness and practicality for medical image segmentation tasks.

In summary, our contributions are as follows:
\begin{itemize}
    \item We propose Vision-FAN, a vision backbone network integrated with FAN layers and self-attention mechanisms. By incorporating Fourier analysis, the model effectively captures long-range dependencies and periodic relationships. This hybrid design addresses the challenges of long-range dependency and periodicity modeling, while significantly enhancing the model's generalization ability across diverse data distributions.
    \item We propose FAN-UNet, marking the first exploration of FAN-based models for potential applications in medical image segmentation.
    \item We conducted extensive performance evaluations of the proposed model. The results demonstrate that our model achieves high accuracy (96.07\%), mIoU (78.83\%), and DSC (88.16\%), validating its effectiveness.
\end{itemize}

\section{Preliminaries}
Fourier analysis, by decomposing a function into its constituent frequencies, reveals the underlying periodic structures within complex functions. At the core of this analysis lies the Fourier series, which represents a periodic function as an infinite sum of sine and cosine terms. Mathematically, the Fourier series expansion for a function f(x) can be expressed as:
\begin{equation}
    f(x)=a_0+\sum_{n=1}^\infty\left(a_n\cos\left(\frac{2\pi nx}{T}\right)+b_n\sin\left(\frac{2\pi nx}{T}\right)\right),
\end{equation}
where, \( T \) represents the period of the function, and the coefficients \( a_n \) and \( b_n \) are determined by integrating the function over one period:
\begin{equation}
    a_n=\frac{1}{T}\int_0^Tf(x)\cos\left(\frac{2\pi nx}{T}\right) dx,\quad b_n=\frac{1}{T}\int_0^Tf(x)\sin\left(\frac{2\pi nx}{T}\right) dx.
\end{equation}
The power of the Fourier series lies in its ability to represent a wide variety of functions, including non-periodic functions through periodic extension, thereby extracting frequency components effectively. Building upon this mathematical foundation, Fourier Analysis Layers aim to embed periodic features directly into network architectures, enhancing generalization and performance across various tasks, especially in scenarios requiring the recognition of patterns and regularities.

\section{Method}
\begin{figure}
    \centering
    \includegraphics[width=0.9\linewidth]{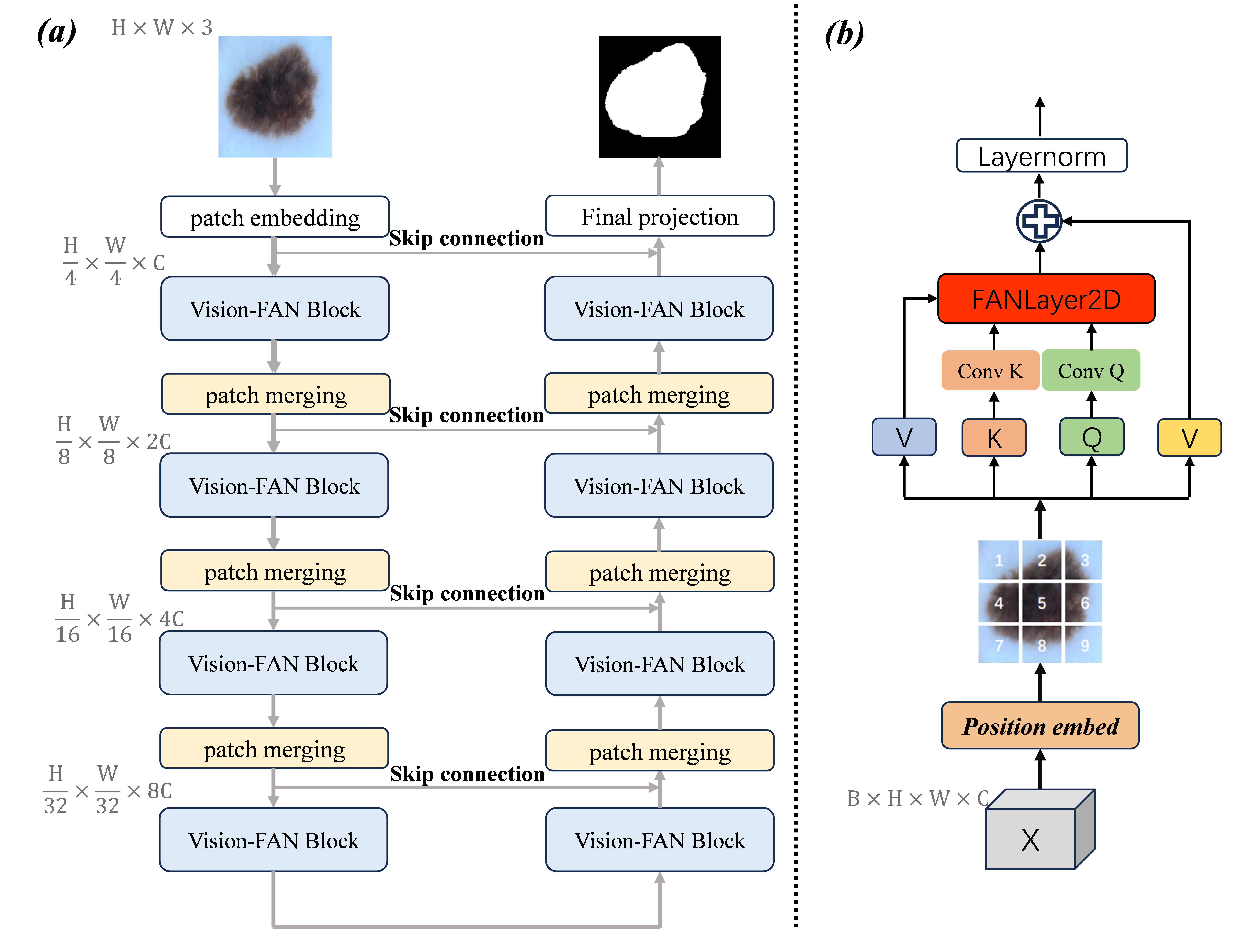}
    \caption{\textbf{(A) }Overall structure of FAN-Unet. \textbf{(B)} Overall structure of Vision-FAN Block}
    \label{fig:model-structure}
\end{figure}

The structure of FAN-Unet is shown in \Cref{fig:model-structure}(a). FAN-Unet aims to  achieve more accurate medical image segmentation by combining the feature extraction capabilities of the U-Net architecture with the Vision-FAN Block based on FAN. 

\subsection{Vision-FAN Block}
Vision-FAN Block is the core module of FAN-UNet, consisting of FANLayer2D and attention mechanisms, effectively modeling periodic features and long-range dependencies, as illustrated in \Cref{fig:model-structure}(b). Specifically, Vision-FAN Block is integrated into the U-Net structure to process 2D features, capturing global long-range dependencies and periodic relationships in images. The design combines position encoding, window-based self-attention, and FANLayer2D, leveraging the strengths of these components to significantly enhance the model’s representation capacity.

First, Vision-FAN Block applies position encoding to the downsampled feature maps, enabling the model to understand the relative relationships of each position in the input feature map. Subsequently, window-based self-attention is employed to capture global spatial dependencies. The self-attention mechanism calculates long-range interactions between different positions, allowing the model to comprehensively understand the global features of the input image. This is particularly critical in medical imaging, as lesion regions often span across different spatial locations. Thus, the introduction of position encoding and window-based self-attention ensures efficient computation while capturing global dependencies.

Next, the output of the self-attention mechanism is passed to FANLayer2D, which serves as a feedforward network to further enhance the modeling of periodic and nonlinear features. FANLayer2D, based on Fourier analysis, explicitly encodes the periodic patterns of the input features, enabling the model to directly capture repetitive structures and textures commonly found in medical images. This periodicity modeling is especially well-suited for medical image analysis, as tissue and pathological features in medical images often exhibit regularity and consistency. By employing Fourier activation functions (including sine and cosine functions) and nonlinear activation functions, FANLayer2D effectively captures the complex patterns and periodic characteristics of the input data.

To ensure a stable training process, Vision-FAN Block incorporates residual connections and layer normalization. Residual connections allow the model to retain input features during learning, reducing reliance on erroneous dependencies from deeper layers. Layer normalization ensures consistent feature distributions, mitigating gradient vanishing or exploding problems, accelerating convergence, and improving model stability.

By integrating Vision-FAN Block into FAN-UNet, the model can simultaneously capture global long-range dependencies and explicitly learn the periodic components of the input data. This combination effectively enhances the model’s capability to identify complex structures and texture patterns in medical images. For instance, in skin lesion segmentation tasks, the shapes and sizes of lesion boundaries and regions often exhibit strong spatial consistency and regularity. Vision-FAN Block leverages these properties to significantly improve segmentation accuracy.

\subsection{FANLayer2D}
The FANLayer2D is the core module of Vision-FAN, specifically designed for handling 2D image data to capture periodic patterns within input features. To understand the construction of FANLayer2D, consider a network layer $f_S(x)$ that represents the Fourier series expansion $F\{f(x)\}$. According to the Fourier series expansion, it can be expressed as:
\begin{equation}
    f_S(x)\triangleq a_0+\sum_{n=1}^N\left(a_n\cos\left(\frac{2\pi nx}T\right)+b_n\sin\left(\frac{2\pi nx}T\right)\right)
\end{equation}
This can be further simplified as:
\begin{equation}
f_S(x)=a_0+\sum_{n=1}^N\left(w_n^c\cos\left(w_n^ix\right)+w_n^s\sin\left(w_n^ix\right)\right)
\end{equation}
and ultimately written in matrix form as:
\begin{equation}
    f_S(x)=B+W^c\cos(W^{in}x)+W^s\sin(W^{in}x)
\end{equation}
where $B\in R^{d_y},W^{in}\in R^{N\times d_x},W^c,W^s\in R^{d_y\times N}$ are learnable parameters. Based on this Fourier series expansion, the FAN layer explicitly incorporates the principles of Fourier transformations, enabling the network to capture periodic components within input features.

Therefore, the design of the FAN layer follows two key principles:1) The expressiveness of Fourier coefficients should scale with depth: Each layer of the network must have sufficient capacity to express Fourier coefficients, ensuring deeper layers possess enhanced periodic modeling abilities; 2) The output of any hidden layer should be capable of modeling periodicity through subsequent layers: This ensures that periodic features are effectively captured and propagated at every level, enhancing the overall expressiveness of the model.

Based on these principles, the FAN layer can be defined as:
\begin{equation}
    \phi(x)\triangleq[\cos(W_px)||\sin(W_px)||\sigma(B_p+W_p^{\prime}x)]
\end{equation}
where $W_p\in R^{d_x\times d_p},W_p^{\prime}\in R^{d_x\times d_{p^{\prime}}},B_p\in R^{d_{p^{\prime}}}$ are learnable parameters.

\begin{figure}
    \centering
    \includegraphics[width=0.9\linewidth]{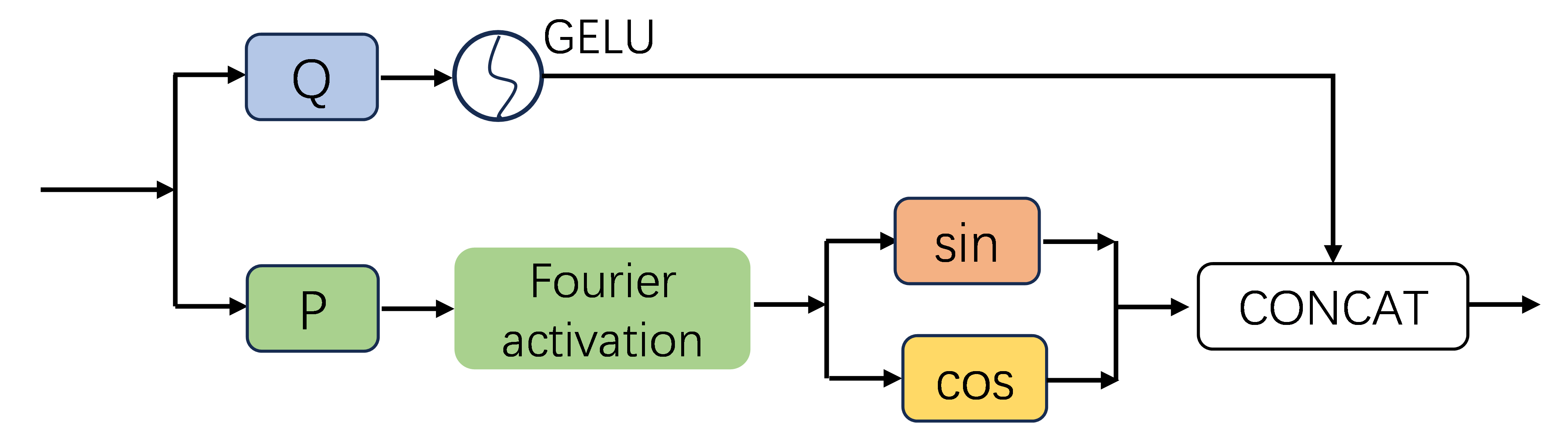}
    \caption{Schematic diagram of FANLayer2D}
    \label{fig:FANLayer2D}
\end{figure}

FANLayer2D is an extension of the above FAN concept to 2D, specifically designed for processing 2D image features, as illustrated in \Cref{fig:FANLayer2D}. For an input feature map $\mathbf{X}\in\mathbb{R}^{B\times C_{\text{in}}\times H\times W}$,  where B is the batch size, $C_in$ is the number of input channels, and H,W are height and width, FANLayer2D first Compute the periodic component P and the nonlinear component G:
\begin{equation}
    \mathbf{P=W_p*X+b_p,}\quad\mathbf{G=W_g*X+b_g}
\end{equation}
Then, apply a nonlinear activation function to G:
\begin{equation}
    \mathbf{G}_{\mathrm{act}}=\phi(\mathbf{G})
\end{equation}
Next, Apply Fourier activation functions to P to obtain periodic features:
\begin{equation}
    \mathbf{P}_{\cos}=\cos(\mathbf{P}),\quad\mathbf{P}_{\sin}=\sin(\mathbf{P})
\end{equation}
Finally, concatenate the periodic features and activated nonlinear features along the channel dimension to form the output features.

Building on this structure, FANLayer2D effectively captures both the periodic and nonlinear characteristics of input data in the 2D space. Its design enables the network to directly learn and represent repetitive patterns and intricate textures present in images, significantly enhancing its capability to accurately segment lesion areas in medical imaging.

A key advantage of FANLayer2D lies in its ability to explicitly encode periodic information, a critical aspect in many medical imaging tasks. For instance, the features of tissues and lesions in medical images often demonstrate a high degree of regularity and consistency. By leveraging Fourier activation functions, FANLayer2D efficiently captures these periodic patterns, enabling the model to more effectively identify and segment these regions with greater precision.

\subsection{Loss function}
FAN-Unet is proposed as a means of obtaining more stable results and higher accuracy in medical image segmentation tasks. Therefore, we employ the Diceloss and Cross-entropy loss function, which are the most elementary in medical image analysis. We then combine these two elements. By calculating the loss at the batch level, we have devised a loss function that can mitigate the fluctuations in loss observed for individual samples, which may be attributable to random noise or misclassification of minor structures. This enables the model to converge in a more stable manner. Furthermore, it can enhance the overall segmentation performance, particularly in scenarios where data distribution is uneven, and facilitate more accurate weighting of different samples. The specific loss function is presented in \cref{equ:loss}.$\alpha$ refer to the weights of loss functions, which is set to $0.5$ by default.
\begin{equation}
\label{equ:loss}
\left\{
\begin{aligned}
\text{Loss}_{\text{B}} &= (1-\alpha)\times\text{CE}_{\text{B}}+\alpha\times\text{Dice Loss}_{\text{B}} \\
\mathrm{CE}_{\mathrm{B}} &= -\frac{1}{B}\sum_{b=1}^{B}\text{Target}_{b}\log(\text{Input}_{b}) \\
\text{Dice loss}_{\text{B}} &= 1-\frac{2\sum_{b=1}^{B}\mathrm{Input}_{b}^{}\times\mathrm{Target}_{b}^{}}{\sum_{b=1}^{B}\left(\mathrm{Input}_{b}^{}+\mathrm{Target}_{b}^{}\right)+\epsilon}
\end{aligned}
\right.
\end{equation}

\section{Experiment}
\subsection{Datasets}
We conduct comprehensive experiments on FAN-Unet for medical image segmentation tasks. Specifically, we evaluate the performance of FAN-Unet on medical image segmentation tasks on the ISIC17\cite{codella2018skin} and ISIC18\cite{codella2019skin,tschandl2018ham10000} datasets.

\begin{itemize}
    \item \textbf{ISIC2017}:The ISIC2017 dataset contains three categories of diseases, melanoma, seborrheic keratosis, and benign nevus, 2,750 images, ground truth, and category labels. There are 2,000 images in the training set, 150 images in the validation set, and 600 images in the test set, and the color depth of the skin disease images is 24 bits, and the image sizes range from 767×576 to 6,621×4,441. The validation and test sets also include unlabeled hyperpixel images. The category labels are stored in tables and the datasets need to be preprocessed before training the model.
    
    \item \textbf{ISIC2018}:The ISIC2018 dataset contains different numbers of disease images for classification and segmentation, for the segmentation task, a total of 2,594 images were used as the training set, and 100 and 1,000 images were used as the validation and test sets, respectively. For the classification task, a total of 12,500 images were included, of which the training set contained a total of 10,015 images of 7 categories of diseases, namely actinic keratoses (327), basal cell carcinoma (514), benign keratoses (1,099), dermatofibromas (115), melanomas (1,113), melanocytic naevi (6,705), and vascular skin lesions (142). The seven classes of images in the classification task dataset are mixed in the same folder, and the labels are stored in tables that require preprocessing.
\end{itemize}

\subsection{Comparison with SOTA models}
We compare FAN-Unet with some state-of-the-art models and some recent mamba-based model, presenting the experimental results in \Cref{tab:isic}. 

For the ISIC2017 and ISIC2018 datasets, FAN-Unet performs well on mIoU and Dice compared to other models. Specifically, FAN-Unet has a 1.11\% and 0.87\% advantage over HC-Mamba on mIoU and Dice, respectively, while it has a 2.01\% and 2.26\% advantage over Unet on mIoU and Dice.
\begin{table}[!th]
	\setlength\tabcolsep{3pt}
	\renewcommand\arraystretch{1.25}
	\caption{Comparative experimental results on the ISIC dataset. (\textbf{Bold} indicates the best.)}
	\begin{center}
		\begin{tabular}{c|c|ccccc}
			\hline
			\textbf{Dataset} &\textbf{Model}          & \textbf{mIoU(\%)$\uparrow$}  & \textbf{DSC(\%)$\uparrow$}   & \textbf{Acc(\%)$\uparrow$}   & \textbf{Spe(\%)$\uparrow$}   & \textbf{Sen(\%)$\uparrow$}   \\ \hline
			\multirow{5}{*}{ISIC17} &UNet\cite{ronneberger2015u}  &76.98 &85.99 &94.65 &97.43 & 86.82  \\
			&UTNetV2\cite{utnetv2}                & 76.35          & 86.23          & 94.84          & 98.05          & 84.85          \\
			&TransFuse\cite{zhang2021transfuse}               & 77.21          & 86.40          & 95.17          & 97.98          & 86.14 \\
            &MALUNet\cite{malunet}  &76.78 &87.13 &95.18 &\textbf{98.47} &84.78 \\
			&VM-UNet\cite{ruan2024vmunet} &77.59 &87.03 & 95.40 &97.47 &86.13   \\
                &HC-Mamba\cite{xu2024hc} & 77.88 &87.38 &95.17 &97.47 &86.99 \\
                 &Med-TTT\cite{xu2024med} & 78.83 & 88.16 & 96.07& 97.86& 87.21\\
                 &FAN-Unet & \textbf{78.99}& \textbf{88.25}& \textbf{96.30}& 97.98 & \textbf{87.91}\\
                 \hline\hline
			\multirow{8}{*}{ISIC18}&UNet\cite{ronneberger2015u}                     & 77.86          & 87.55          & 94.05          & 96.69          & 85.86          \\
			&UNet++ \cite{unet++}                & 76.31          & 85.83          & 94.02          & 95.75          & 88.65          \\
			&Att-UNet \cite{attentionunet}             & 76.43          & 86.91          & 94.13          & 96.23          & 87.60          \\    
			&SANet \cite{sanet}                 & 77.52          & 86.59          & 93.39          & 95.97          & \textbf{89.46} \\
			&VM-UNet\cite{ruan2024vmunet} &77.95 &87.61  &94.13   &96.94&85.23 \\
            &HC-Mamba\cite{xu2024hc} &78.42 & 87.89 & 94.24 & 96.98 & 88.90 \\ 
            &Med-TTT\cite{xu2024med} & 78.59 & 88.01 & 94.30 & \textbf{96.98} & 85.95\\
            &FAN-Unet & \textbf{78.74}& \textbf{88.11}& \textbf{94.31}& 96.79 & 86.60\\
            \hline
		\end{tabular}
		\label{tab:isic}
	\end{center}
\end{table}
\subsection{Ablation experiments}
In this section, we demonstrated the effectiveness of Positional Embed, self-attention mechanism, and Vision-FAN Block through ablation experiments. We constructed four network configurations for this purpose, which are listed in the following. (1) Proposed w/o Vision-FAN Block: replace Vision-FAN Block with normal convolution. (2) Proposed w/o Positional Embed: incorporates Vision-FAN Block but without Positional Embed. (3) Proposed w/o self-attention mechanism: incorporates Vision-FAN Block but without self-attention mechanism. (4) Proposed: the full proposed architecture.

\Cref{tab:abl} summarizes the ablation experimental results, further emphasizing the superiority of the proposed network across multiple metrics, including mIoU, Dice, accuracy (Acc), and sensitivity (Sen), while ranking second in specificity (Spe). In comparison, the 'Proposed w/o Positional Embed' configuration achieved moderate performance, whereas the 'Proposed w/o Vision-FAN Block' configuration performed poorly in the quantitative experiments. These objective evaluations demonstrate the effectiveness of the Vision-FAN Block and Positional Embed.

\begin{table*}[!htbp]
\centering
\caption{Results of the ablation experiments}
\label{tab:abl}
\begin{tabular}{lccccc}
\hline
\textbf{Configuration} & \textbf{mIoU(\%)} & \textbf{Dice(\%)} & \textbf{Acc(\%)} & \textbf{Spe(\%)} & \textbf{Sen(\%)} \\
\hline
Proposed w/o Vision-FAN Block & 77.86 & 87.55 & 94.05 & 96.69 & 85.86 \\
Proposed w/o Positional Embed & \underline{78.42} & 87.89 & \underline{94.21} & \textbf{96.94} & \underline{86.23} \\
Proposed w/o self-attention mechanism & 78.22 & \underline{88.01} & 93.29 & 96.02 & 86.11 \\
\textbf{Proposed} & \textbf{78.74}& \textbf{88.11}& \textbf{94.31} & \underline{96.79} & \textbf{86.60} \\
\hline
\multicolumn{6}{l}{\textbf{Bold} indicates the best, the second-best values are \underline{underlined}.} \\
\end{tabular}
\end{table*}

\section{Conclusion}
We propose a novel medical image segmentation model, FAN-UNet, which integrates the Vision-FAN Block, U-Net architecture, and positional encoding to address the limitations of existing CNN- and Transformer-based models. The Vision-FAN Block directly models periodic features from data, effectively capturing both long-range dependencies and periodic relationships. Furthermore, the combination of positional encoding and the U-Net structure enables efficient feature extraction from medical images, while accurately capturing relative positional relationships, thereby enhancing the modeling of periodic features.

Extensive experiments conducted on multiple datasets demonstrate that FAN-UNet achieves outstanding performance in terms of accuracy, mIoU, and DSC, particularly in challenging segmentation scenarios. These results validate the effectiveness and robustness of the proposed model, making it a valuable contribution to the field of medical image analysis. Future work will explore the application of FAN-UNet to other medical imaging modalities and further enhance its generalization capabilities.

\bibliographystyle{unsrt}  
\bibliography{references}  

\end{document}